\RequirePackage{etoolbox}
\csdef{input@path}%
{%
 {sty/},
 {img/},
}

\documentclass{elsevierbook}

\usepackage{booktabs} 
\usepackage[ruled]{algorithm2e} 

\usepackage{pdflscape}
\usepackage{rotating}
\usepackage{enumitem}
\usepackage{color,soul}

\SetAlFnt{\small}
\SetAlCapFnt{\small}
\SetAlCapNameFnt{\small}
\SetAlCapHSkip{0pt}
\IncMargin{-\parindent}

\let\Bbbk\relax
\usepackage{amssymb}

\usepackage{pifont}

\newcommand{\eg}{{\it e.g., }}
\newcommand{\etal}{{\it et~al. }}
\newcommand{\ie}{{\it i.e., }}

\usepackage{graphicx}
\usepackage{caption}
\usepackage{subcaption}

\usepackage{enumitem}
\begin{document}

\tableofcontents
\clearpage

\newlist{abbrv}{itemize}{1}
\setlist[abbrv,1]{label=,labelwidth=1in,align=parleft,itemsep=0.1\baselineskip,leftmargin=!}
 
\chapter*{List of Abbreviations}
 
\begin{abbrv}
 
\item[VOD]			Video On Demand
\item[DRM]			Digital Rights Management 
\item[OTT]			Over The Top
\item[QoE]			Quality of Experience
\item[AWS]			Amazon Web Services
\item[VM]				Virtual Machine
\item[GOP]			Group Of Pictures
\item[MB]				Macroblock
\item[MPEG]			Moving Picture Experts Group
\item[AVC]			Advanced Video Coding 
\item[VP]				Video Phone
\item[HEVC]			High Efficiency Video Coding
\item[ISOBMFF]		ISO Base Media File Format
\item[HTTP]			HyperText Transfer Protocol
\item[RTMP]			Real-Time Messaging Protocol
\item[RTSP]			The Real Time Streaming Protocol
\item[HLS]				HTTP Live Streaming
\item[DASH]			Dynamic Adaptive Streaming over HTTP
\item[CDN]			Content Delivery Networks
\item[P2P]				Peer-to-Peer
\item [MCU]			Multipoint Control Unit
\item[DVD]			Digital Versatile Disc
\item[UDP]			User Datagram Protocol
\item[RTP]			Real-Time Protocol
\item[MPD]			Media Presentation Description
\item[XML]			Extensible Markup Language
\item[CMAF]			Common Media Application Format
\item[M2TS]			MPEG-2 Transport Stream
\item[M3U8]			MP3 Playlist File (UTF-8)
\item[SDTP]			Stall Duration Tail Probability
\item[CPU]			Central Processing Unit
\item[VRC]			Video Cassette Recording
\item[MAC]			Message Authentication Code
\item[CVSE]			Cloud-based Video Streaming Engine
\item[SLA]				Service Level Agreement
\item[SLO]			Service Level Objectives

\end{abbrv}
\newpage
\chapter[A Survey on Cloud-Based Video Streaming Services]{A Survey on Cloud-Based Video Streaming Services} 
\subchapter{Xiangbo Li, Mahmoud Darwich, Magdy Bayoumi, Mohsen Amini Salehi}

  \begin{abstract}
Video streaming, in various forms of video on demand (VOD), live, and 360 degree streaming, has grown dramatically during the past few years. In comparison to traditional cable broadcasters whose contents can only be watched on TVs, video streaming is ubiquitous and viewers can flexibly watch the video contents on various devices, ranging from smart-phones to laptops and large TV screens. Such ubiquity and flexibility are enabled by interweaving multiple technologies, such as video compression, cloud computing, content delivery networks, and several other technologies. 
As video streaming gains more popularity and dominates the Internet traffic, it is essential to understand the way it operates and the interplay of different technologies involved in it. Accordingly, the first goal of this paper is to unveil sophisticated processes to deliver a raw captured video to viewers' devices. In particular, we elaborate on the video encoding, transcoding, packaging, encryption, and delivery processes. We survey recent efforts in academia and industry to enhance these processes. As video streaming industry is increasingly becoming reliant on cloud computing, the second goal of this survey is to explore and survey the ways cloud services are utilized to enable video streaming services. The third goal of the study is to position the undertaken research works in cloud-based video streaming and identify challenges that need to be obviated in future to advance cloud-based video streaming industry to a more flexible and user-centric service.  
\end{abstract}




\section{Introduction}\label{sec:intro}
\subsection{Overview}
The idea of receiving a stream of video contents dates back to the invention of television in the early years of the 20th century. However, the medium on which people receive and watch video contents has substantially changed during the past decade---from  conventional televisions to streaming on a wide variety of devices (\eg laptops, desktops, and tablets) via Internet. Adoption of the Internet-based video streaming is skyrocketing to the extent that it has dominated the whole Internet traffic. A report by  Global Internet Phenomena shows that video streaming has already accounted for more than 60\% of the whole Internet traffic~\cite{intro_1}. The number of Netflix\footnote{https://www.netflix.com} subscribers has already surpassed cable-TV subscribers in the U.S.~\cite{netflixtv}. 

Nowadays, many Internet-based applications function based on video streaming. Such applications include user-generated video contents (\eg those in YouTube\footnote{https://www.youtube.com}, Vimeo\footnote{https://www.vimeo.com}), live streaming and personal broadcasting through social networks (\eg UStream\footnote{https://video.ibm.com} and Facebook Live\footnote{https://www.facebook.com}), over the top (OTT) streaming (\eg Netflix and Amazon Prime\footnote{https://www.amazon.com}), e-learning systems~\cite{elearningvideo} (\eg Udemy\footnote{https://www.udemy.com}), live game streaming platform (\eg Twitch\footnote{https://www.twitch.tv}), video chat and conferencing systems~\cite{lin2003dynamic}, natural disaster management and security systems that operate based on video surveillance~\cite{collins2000system}, and network-based broadcasting channels (\eg news and other TV channels)~\cite{joo2008effective}.

As video streaming services grow in popularity, they demand more computing services for streaming. The uprising popularity and adoption of streaming has coincided with the prevalence of cloud computing technology. Cloud providers offer a wide range of computing services and enable users to outsource their computing demands. Cloud providers relieve video streaming providers from the burden and implications of maintaining and upgrading expensive computing infrastructure~\cite{li2018cost}. Currently, video stream providers are extensively reliant on cloud services for most or all of their computing demands~\cite{he2013toward}. The marriage of video streaming and cloud services has given birth to a set of new challenges, techniques, and technologies in the computing industry. 

Although numerous research works have been undertaken on cloud-based video streaming, to our knowledge, there is no comprehensive survey that shed lights on challenges, techniques, and technologies in cloud-based video streaming. As such, the essence of this study is to \emph{first}, shed light on the sophisticated processes required for Internet-based video streaming; \emph{second}, provide a holistic view on the ways cloud services can aid video stream providers; \emph{third}, provide a comprehensive survey on the research studies that were undertaken in the intersection of video streaming and cloud computing; and \emph{fourth} discuss the future of cloud-based video streaming technology and identify possible avenues that require further research efforts from industry and academia.

Accordingly, in this study, we first explain the way video streaming works and elaborate on each process involved in it. Then, in Section~\ref{sec:issues}, we provide a holistic view of the challenges and demands of the current video streaming industry. Next, in Section~\ref{sec:cbvs}, we discuss how cloud services can fulfill the demands of video streaming, and the survey the research works undertaken for that purpose. In the end, in Section~\ref{sec:conclusion}, we discuss the emerging research areas in the intersection of video streaming and cloud computing.

\subsection{Cloud Computing for Video Streaming}

To provide a high Quality of Experience (QoE) for numerous viewers scattered worldwide with diverse display devices and network characteristics, video stream providers commonly pre-process (\eg pre-transcode and pre-package) and store the video contents of multiple versions~\cite{pre_3}. As such, viewers with different display devices can readily find the version matches their devices. An alternative approach to pre-processing videos is to process them in a lazy (\ie on-demand) manner~\cite{pre_3, li2018cost}. In particular, this approach can be useful for videos that are rarely accessed. Recent analytical studies on the statistical patterns of accessing video streams (\eg~\cite{intro_9}) reveal that the videos of a repository are not uniformly accessed. In fact, streaming videos follows a long-tail pattern~\cite{miranda2013characterizing}. That means, many video streams are rarely accessed and only a small portion (approximately 5\%) of the videos (generally referred to as \emph{hot video streams}) are frequently accessed.   
Both pre-processing and on-demand approaches need the video streaming provider to provide an enormous computing facility. 

Maintaining and upgrading in-house infrastructure to fulfill the computational, storage, and networking demands of video stream providers is costly. Besides, it is technically far from the mainstream business of stream providers, which is video content production and publishing. Alternatively, cloud service providers, such as Amazon Cloud (AWS), Google Cloud, and Microsoft Azure, can satisfy these demands by offering efficient services with a high availability~\cite{pre_5}. Video stream providers have become extensively reliant on cloud services for most or all of their computing demands. For instance, Netflix has outsourced the entirety of its computational demands to Amazon cloud~\cite{netflixamazon}. 

In spite of numerous advantages, deploying cloud services has presented new challenges to video stream providers. In particular, as cloud providers charge their users in a pay-as-you-go manner for their services~\cite{ica3pp10}, the challenge for stream providers is to minimize their cloud expenditure, while offering a certain level of QoE for their viewers. 

Numerous research works have been conducted to overcome the challenges of video streaming using cloud services. For instance, researchers have studied the cost of deploying cloud services to transcode videos~\cite{deneke2014video,pre_5}, the cost-benefit of various video segmentation models~\cite{bg_2, pre_3}, applying customized scheduling methods for video streaming tasks~\cite{rw_11,deneke2014video}, and resource (Virtual Machine) provisioning methods for video streaming~\cite{pre_3, rw_10}. Nonetheless, these studies mostly concentrate on one aspect of video streaming processing and how that aspect can be performed efficiently on the cloud. Further studies are required to position these works in a bigger picture and provide a higher level of vision on the efficient use of cloud services for the entire workflow of video streaming.


\section{The Mystery of Video Streaming Operation}\label{sec:bkgnd}
\subsection{Structure of a Video Stream}

\begin{figure}[htb] 
    \centering
    \includegraphics[width=2.5in]{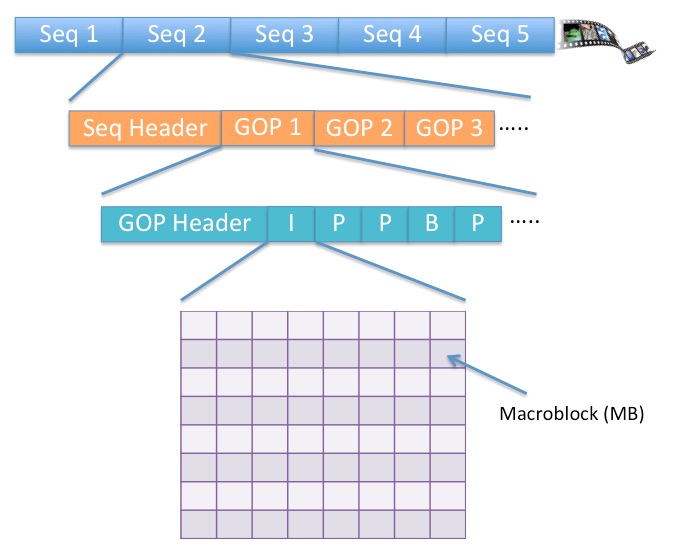}
    \caption{The structural overview of a video stream. A video stream contains multiple segments and each segment contains multiple GOPs. A frame in a GOP includes a number of macroblocks (MB).}
    \label{video_segment}
\end{figure}

As shown in Figure~\ref{video_segment}, a video stream consists of a sequence of multiple smaller segments. Each segment contains several \textit{Group Of Pictures} (GOP) with a segment header at the beginning of each GOP. The segment header includes information such as the number of GOPs in that segment and the type of the GOPs.
A GOP is composed of a sequence of frames. The first frame is an \texttt{I} (intra) frame, followed by several \texttt{P} (predicted) and \texttt{B} (bi-directional predicted) frames.
A frame in a GOP is divided into multiple \textit{slices} and each slice consists of several \textit{macroblocks} (MB). The MBs are considered as the unit for video encoding and decoding operations. 

Two types of GOP can exist, namely closed-GOP and open-GOP. In the former, GOPs are independent, \ie there is no relation between GOPs. Therefore, closed-GOPs can be processed independently. Alternatively, in open-GOP, a GOP is dependent on another GOP, hence, cannot be processed independently. 

To process video streams, they can be split at different levels, namely segment level, GOP level, frame level, slice level, and macroblock level. Sequence-level contains several GOPs that can be processed independently. However, due to the large size of a sequence, its transmission and processing time become a bottleneck~\cite{bg_14}. Processing at the frame, slice, and macroblock levels implies dealing with the spatio-temporal dependencies that makes the processing complicated and slow~\cite{bg_14}. In practice, video stream providers generally perform processing at the segment or GOP levels. That is, they define a segment or a GOP as a unit of processing (\ie a \emph{task}) that can be processed independently~\cite{bg_2}.

\subsection{Video Streaming Operation Workflow}\label{subsec:workflo}

In both Video On Demand (VOD) (\eg Hulu, YouTube, Netflix) and Live streaming (\eg Livestream\footnote{https://livestream.com/}), the video contents generated by cameras have to go through a complex workflow of processes before being played on viewers' devices. In this section, we describe these processes.

\begin{landscape}
\begin{figure}
    \centering
    \vspace*{3cm}
    \includegraphics[width=8in]{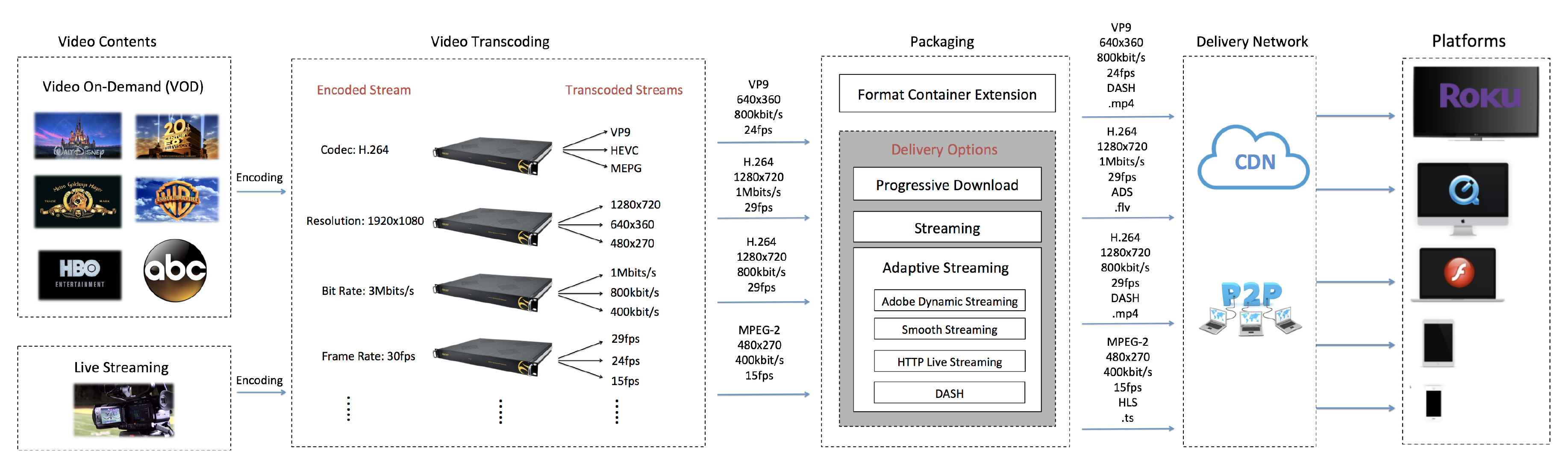}
    \caption{A bird's-eye view to the workflow of processes performed on a video stream from the source to the viewer's device. The workflow includes compression, transcoding, packaging, and delivery processes.}
    \label{fig:achitecture}
\end{figure}
\end{landscape}

Figure~\ref{fig:achitecture} provides a bird's-eye view of the main processes performed for streaming a video---from video production to playing the video on viewers' devices. These processes collectively enable raw and bulky videos, generated by cameras, be played on a wide variety of viewers' devices in a real-time manner and with the minimum delay. It is noteworthy that, in addition to these processes, there are generally other processes to enable features such as video content protection and cost-efficiency of video streaming. In the rest of this section, we elaborate on the main processes required for video streaming. Additional processes (\eg for video content protection and analysis of video access rates) are discussed in the later parts of the paper (Sections~\ref{sec:security},~\ref{sec:analysis}, and~\ref{sec:storage}, respectively). 

The first step in video streaming is video content production. A raw video content generated by cameras can consume enormous storage space, which is impossible to be transmitted via current Internet speed. For instance, one second of raw video with 4K resolution occupies approximately 1 GB storage. Therefore, the generated video, firstly, has to be compressed, which is also known as \emph{video encoding}. The concept of the video is just continuously showing a large number of frames at a certain rate (aka frame rate) to create a moving delusion. This large number of frames usually contains spatial and temporal redundancies both inside a frame and between successive frames. In the video compression process, these redundancies are removed using a certain compression standard, \eg H.264/MPEG-4 AVC~\cite{schwarz2007overview}, VP9~\cite{mukherjee2013latest}, and H.265/HEVC~\cite{intro_3}. The compression process encodes the raw video based on a specific compressing standard (known as \emph{codec}), resolution, bit rate, and frame rate at a significantly smaller size.

A viewer's display device generally can decode videos with a certain compression format. Therefore, to support heterogeneous display devices, a video encoded in a certain format has to be converted (also called \emph{transcoded}) to various formats. In the transcoding process, a video is first decoded and then encoded to another compression format. Thus, transcoding is generally a computationally intensive process. In Section~\ref{sec:video_transcoding}, we elaborate further on the video transcoding process.

For streaming of a transcoded video to a viewer's device, the video file has to be structured to facilitate transferring and time-based presentation of the video content. Thus, the transcoded video file must be \emph{packaged} based on a certain structure that is supported by the viewer's player. The packaging process is the basis for what is commonly known as the video file format (also, called as \emph{container format}). The container format introduces a header that includes information about the supported streaming protocols and the rules to send video segments.  ISO Base Media File Format (ISOBMFF)~\cite{timmerer2010http}, which is the basis for \emph{MP4} format, and 3GPP TS~\cite{transport_ts}, which is the basis for \emph{ts} format, are widely used for the packaging process.

To deliver the packaged (\ie formatted) video files over the network, they need to use an application layer network protocols (\eg HTTP~\cite{fielding1999hypertext}, RTMP~\cite{lesser2005programming}, and RTSP~\cite{schulzrinne1998real}). There are various \emph{delivery techniques} that dictates the way a video stream is received and played. Progressive Download~\cite{chen2012progressive}, HTTP Live Streaming (HLS)~\cite{pantos2017http}, and Dynamic Adaptive Streaming over HTTP (DASH)~\cite{stockhammer2011dynamic} are examples of delivery techniques. Each delivery technique is based on a particular network application layer protocol. For instance, HLS and DASH work based on HTTP and Adobe Flash streaming works based on RTMP. Further details about video packaging are discussed in Section~\ref{sec:video_transmuxing}.

Once the video is packaged, it is delivered to viewers around the world through a distribution network. However, due to the Internet transmission delay, viewers located far from video servers and repositories suffer from a long delay to begin streaming~\cite{krishnan2013video}. To reduce this delay, Content Delivery Networks (CDN)~\cite{buyya2008content} are used to cache the frequently watched video contents at geographical locations close to viewers. Another option to reduce the startup delay is to use Peer-to-Peer (\ie serverless) approaches in which each receiver (\eg viewer's computer) acts as a server and sends video contents to another peer viewer~\cite{ramzan2012video}. More details of the delivery network will be discussed in Section~\ref{sec:delivery_content}

\section{Video Streaming: Challenges and Solutions}\label{sec:issues}

\subsection {Overview}
In the previous section, we described the high-level workflow of processes  for video streaming. However, there are various approaches and challenges to accomplish those processes. In this section, we elaborate on all issues, challenges and current solutions that are related to video streaming.

\begin{landscape}
\begin{figure}
    \centering
    \vspace*{2cm}
    \includegraphics[width=8in]{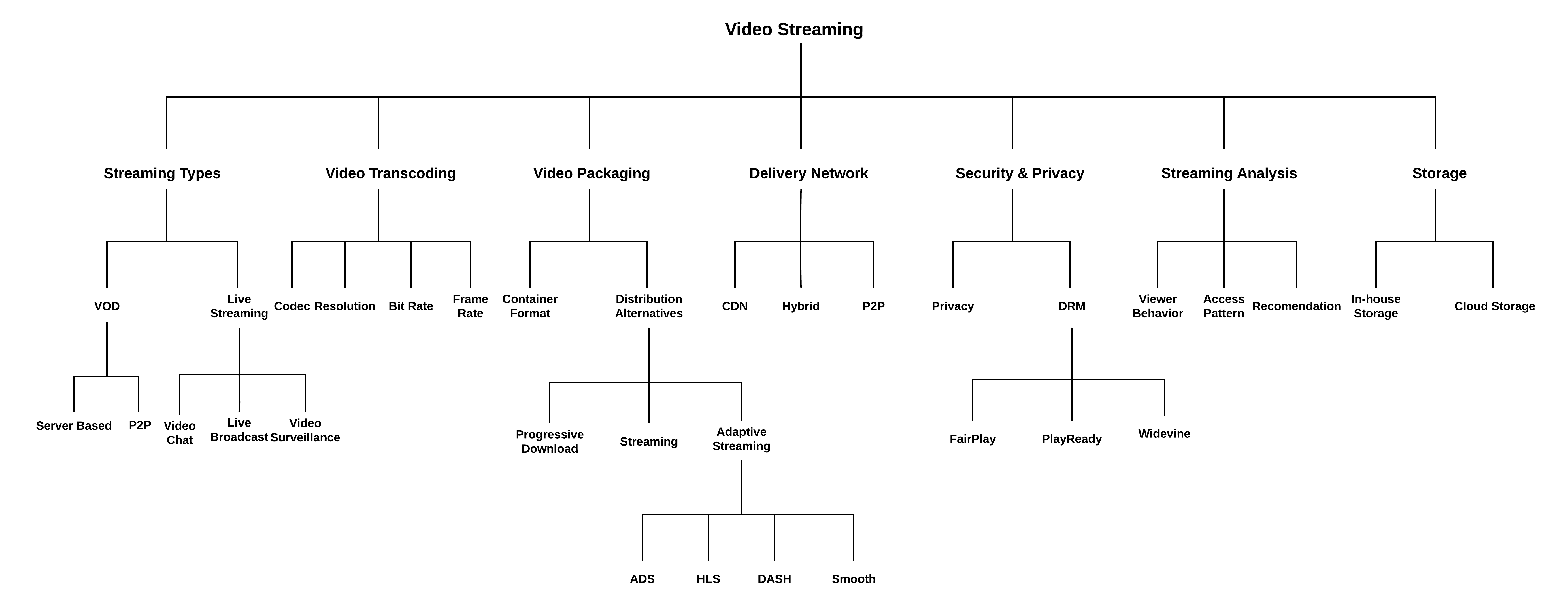}
    \caption{Taxonomy of all aspects we need to deal with in video streaming.}
    \label{fig:tax}
\end{figure}
\end{landscape}

Figure~\ref{fig:tax} presents a taxonomy of all the issues that a video streaming system needs to deal with. The taxonomy shows different types of video streaming (\ie VOD and live streaming). It discusses variations of video transcoding and packaging processes along with important streaming protocols. The taxonomy shows possible ways of distributing video content, such as CDN~\cite{zhuang} and P2P~\cite{liu2008survey}. Video content protection is further detailed on video privacy for video streaming~\cite{zhang2005hiding} and copyright issues, known as Digital Right s Management (DRM)~\cite{drm:content}. The taxonomy covers analytical studies that have been conducted to understand the viewers' behaviors and discovering their access patterns to videos streams. Last, but not the least, the taxonomy covers the storage issue for video streaming and possible strategies for video streaming repository management, such as in-house storage versus outsourcing solutions via cloud storage services. 

\subsection{Video Streaming Types}

\begin{figure}[htbp] 
    \centering
    \includegraphics[width=.8\textwidth]{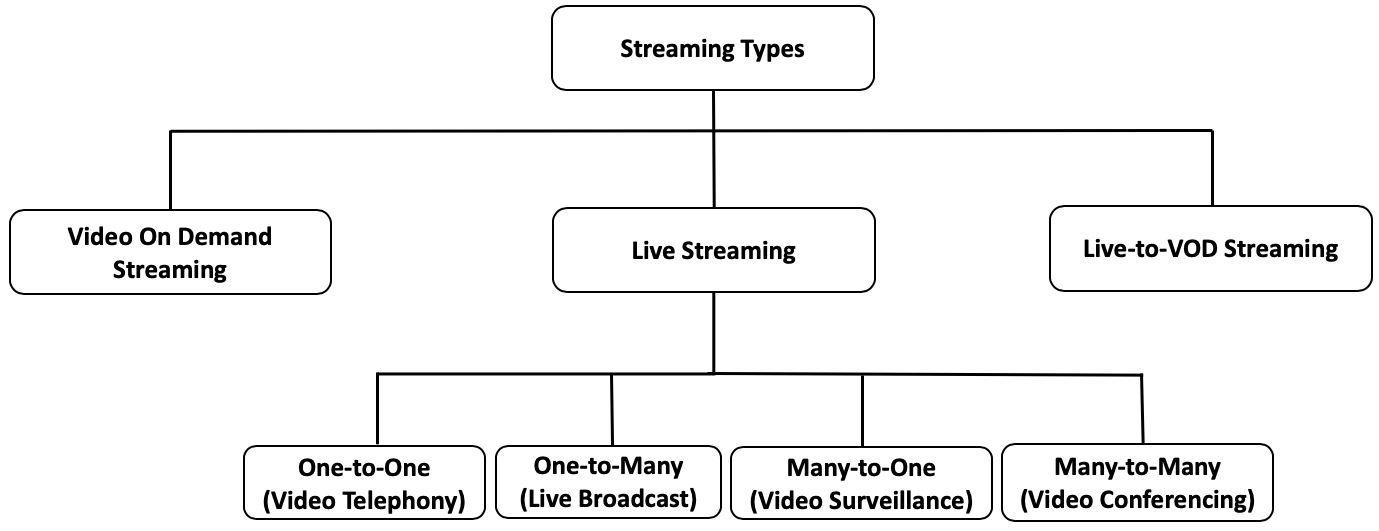}
    \caption{Taxonomy of different types of video streaming.}
    \label{fig:streaming_types}
\end{figure}

The taxonomy shown in Figure~\ref{fig:streaming_types} expresses possible ways a video streaming service can be provided to viewers. More Specifically, video streaming service can be offered in three main fashions: \emph{Video On Demand} (VOD) streaming, \emph{live} streaming, and \emph{live-to-VOD} streaming. 

\subsubsection{VOD Streaming} 
In VOD streaming, which is also known as Over-The-Top (OTT) streaming, the video contents (\ie video files) are already available in a video streaming repository and are streamed to viewers upon their requests. By far, VOD is the most common type of video streaming and is offered by major video streaming providers, such as YouTube, Netflix, and Amazon Prime Video. 

Some VOD providers (\eg Netflix and Hulu) offer professionally made videos and movies that are subscription-based and viewers are generally required to pay a monthly fee to access their service. Alternatively, other VOD services (\eg YouTube) operate based on user-provided videos. Such services are generally advertisement-based and free of charge. VODs have also applications in e-learning systems~\cite{elearningvideo}, Internet television~\cite{iptv14}, and in-flight entertainment systems~\cite{erdemir2017project}.

\subsubsection{Live Streaming} 
In live video streaming, the video contents are streamed to the viewer(s), as they are captured by a camera. Live video streaming has numerous applications, such as event coverage and video calls. The live video streaming used in different applications have minor differences that are mainly attributed to the buffer size on the sender and receiver ends~\cite{stypes:thesis}. In general, a larger buffer size causes a more stable streaming experience but imposes more delay. This delay can be tolerated in live broadcasting applications, however, delay-sensitive applications (\eg video telephony) cannot bear the delay, thus need a shorter buffer size. 
Accordingly, live streaming can have four variations as follows:

\begin{enumerate}
\item[(A)] \textbf{One-to-one} (unicast) streaming is when a user streams video contents to another user. This type is primarily used in video chat and video call applications that require two live streams, one from each participant. This streaming type requires short delays to enable smooth conversation between participants. As such, these applications generally operate with a short buffer size and low picture quality to make the delay as small as possible~\cite{stypes:thesis}. Skype\footnote{https://www.skype.com} and video telephony applications~\cite{jana2014improving} like FaceTime\footnote{https://support.apple.com/en-us/HT204380} are instances of this type of live streaming. 

\item[(B)] \textbf{One-to-many} (multicast) streaming is when one source streams video to many viewers. A well-known example of this type is live broadcasting which is currently offered by many social network applications, such as Facebook and Instagram. Jo \etal \cite{jo2002synchronized} conducted one of the first studies on live streaming. They identified and addressed several challenges in multicast streaming regarding signaling protocols, network stability, and viewer variations.     

\item[(C)]  \textbf{Many-to-one} occurs when several cameras capture scenes and send them to one viewer. The most important application for this type of streaming is multi-camera video surveillance which is used for situational awareness for security purposes or natural disaster management~\cite{hpcc17}. In this type of streaming, the video contents are collected from multiple cameras and displayed on special multi-screen monitoring devices\cite{gualdi2008video}.

\item[(D)] \textbf{Many-to-many} streaming occurs when a group of users in different geographical locations holds a video conference. In this case, all users stream live to all others. For this streaming type, Multipoint Control Unit (MCU)~\cite{stypes:thesis} method can be used to combine individual participants into a single video stream and broadcast it. 
Most of video chat applications, \eg Skype and Google Hangouts, support many-to-many live streaming, in addition to one-to-many streaming.
\end{enumerate}

\subsubsection{Live-to-VOD Streaming}
In addition to live and VOD streaming, we can also consider a combination of live and VOD streaming, known as Live-to-VOD~\cite{livetovod}, as another type of streaming. In this type of streaming, which is mostly used on one-to-many live streaming, the live video stream is recorded and can be readily used in form of VOD streaming. 

Using live-to-VOD streaming, viewers who are not online during the live stream can watch the video at a later time. In addition, live-to-VOD can provide VOD-like services to live stream viewers. For instance, live stream viewers can have the rewind ability. Another application of live-to-VOD is to play live video contents in different time zones. For example, using live-to-VOD, the same TV program that is live streamed at 8:00 am in the Eastern Time Zone, can be played at 8:00 am in the Pacific Time Zone.

\subsubsection{Differences in Processing Live and VOD Streaming}
Although the workflow and processes that are applied to live video streams are the same as those for VOD, there are certain differences between them. Specifically, live and VOD streaming are different on the way they are processed~\cite{pre_5}. 

First, in live streaming, the video segments are processed as they are generated. This has two implications: 
\begin{itemize}
\item The video segments have to be processed (\eg transcoded) on-the-spot, whereas in VOD, it is possible to pre-process videos (\ie in an off-line manner). 

\item There is no historic execution (\ie processing) time information for live video segments~\cite{pre_5}. In contrast, in VOD, each video segment is processed multiple times and the historic execution time information are available. The historic information are particularly important for the efficient scheduling of video streaming tasks.
\end{itemize}

The second difference between live and VOD streaming is the way they are treated for processing. In both live and VOD streaming, to ensure Quality of Experience (QoE), each video streaming task is assigned a deadline that must be respected. That is, the video task processing must be completed before the assigned deadline. The deadline for each video task is determined based on the presentation time of the pertinent video segment. 
If a task cannot meet its assigned deadline for any reason, then, in VOD, the task has to wait until it is processed~\cite{pre_4}. In contrast, in live streaming, if a task misses its deadline, the task must be dropped (\ie discarded) to keep up with the live streaming~\cite{pre_5}. In other words, there is no reason to process a task whose time has passed. 

The third difference between VOD and live streaming is again related to the way deadlines are assigned to video streaming tasks. In fact, in video streaming, if a task misses its deadline, all the tasks behind that, \ie those process video segments later in the stream, should update their streaming deadlines (presentation times) accordingly. This is also known as the dynamic deadline, however, this is not the case in live streaming and the tasks' deadlines in live streaming cannot be changed.

\subsection{Video Transcoding}\label{sec:video_transcoding}

Video contents are originally encoded with one specific spatial resolution, bit rate, frame rate, and compression standard (codec). In order to play the videos on different devices, streaming service providers usually have to adjust the original videos in terms of the viewer's device and network bandwidth. The process of this adjustment is called \textit{video transcoding}~\cite{intro_6, intro_7}. 

Video transcoding is a compute-intensive task and needs powerful computers to process it~\cite{tpds17}. Therefore, video transcoding is generally carried out in an off-line manner, called pre-transcoding~\cite{li2018cost}. In the following subsections, we will elaborate different transcoding operations, respectively.

\subsubsection{Bit Rate}
Video bit rate is the number of bits used to encode a unit time of video. Bit rate directly impacts on video quality, as higher bit rate produces better quality, while higher bit rate consumes larger network bandwidth and storage. In order to stream videos to viewers with different network conditions, streaming service providers usually convert video with multiple bit rates~\cite{bg_3}.

\subsubsection{Spatial Resolution}
Resolution represents the dimensional size of a video, which indicates the number of pixels on each video frame. Therefore, higher resolution contains more pixels and details, as results in larger size. Video resolution usually needs to match with the screen size. Low resolution video plays on large screen will causes blurry after upsample, while high resolution plays on small screen is just a waste of bandwidth since viewer usually won't notice the difference due to the limited pixels on the screen. To adapt to the diverse screen size on the market, original videos have to be transcoded to multiple resolutions~\cite{bg_7}. 
 
\subsubsection{Frame Rate}
When still video frames plays at a certain speed rate, human visual system will feel the object is moving. Frame rate indicates number of video frames shown per second. Videos or films are usually recorded at high frame rate to produce smooth movement, while devices may not support such high frame rate. Therefore, in some cases, videos have to be reduced to a lower frame by removing some frames. On the other hand, increasing frame rate is more complicated than reducing frame rate, since it have add non-existent frames. Overall, to be adaptive to larger scale device, video are transcoded to different frame rates~\cite{bg_12}.

\subsubsection{Video Compression Standard}
Video Compression standard is the key to compress a raw video, the encoding process mainly goes through four steps: prediction, transformation, quantization, and entropy coding, while decoding  is a just reverted encoding process. With different codecs manufactured on different devices (\eg DVD player with MPEG-2~\cite{haskell1996digital}, BluRay player with H.264~\cite{wiegand2003overview}, 4K TV with HEVC~\cite{intro_3}), an encoded video may have to be converted to the supported codec on that device. Changing codec is the most compute-intensive type of transcoding~\cite{tpds17} since it has to decode the bitstream with the old codec first and then encode it again with the new codec.

\subsection{Video Packaging}\label{sec:video_transmuxing}
 Transmitting an encoded/transcoded video file from server to viewer involves multiple layers of network protocols, namely physical layer, data link layer, network layer, transport layer, session layer, presentation layer, and application layer~\cite{jones2001survey}. The protocols of these layers dictate video packaging details, such as stream file header syntax, payload data, authorization, and error handling. Since video streaming protocols operate under the application layer, they potentially can use different protocols in the underlying layers to transmit data packets. 
 
 Choosing the right streaming technology requires understanding pros and cons of the streaming protocols and video packaging (aka container formats). In this section, we discuss three popular streaming technologies plus the streaming protocols and container formats required for each one of the technologies.
  
\subsubsection{Progressive Download}
Back in old days, when online video streaming was not  practical, a video could not be viewed until it was completely downloaded on the  device. This implies that viewers usually had to wait for a considerable amount of time (from minutes to even hours) before watching the video. Progressive download resolved this issue by allowing a video to be played as soon as the player's initial buffer is filled by segments of the video. This reduces the waiting time down to 3--10 seconds to begin watching a video~\cite{balcisoy2004progressive}.

Due to the downloading feature, progressive download can face three problems. First, since a video is downloaded linearly, if the viewer's network bandwidth is too low, the viewer cannot move forward a video until that part is fully downloaded. Second downside of progressive download is that if a video file is fully downloaded, but viewer stops watching in the middle, the rest bandwidths are wasted. Third, copyright is problematic in progressive download because the whole video is downloaded on the viewer's storage device. Progressive Download utilizes HTTP protocol that itself operates based on the TCP  protocol, which provides better reliability and error-resilience than UDP, but it incurs a high network latency~\cite{padmanabhan1995improving}. These inherent drawbacks of HTTP-based progressive download raised the need to a dedicated technology for video streaming.  

\subsubsection{Dedicated Protocol for Video Streaming}

To avoid problems of progressive download, a dedicated protocol for real-time streaming (known as RTP)~\cite{jacobson2003rtp} was created. This protocol delivers video contents from a separated streaming server. While traditional HTTP servers handle web requests, streaming servers only handle streaming. The connection is initiated between the player and the streaming server whenever a viewer clicks on a video in a web page. The connection persists until the video terminates or the viewer stops watching it. In comparison to stateless HTTP, RTP protocol is considered stateful because of this persistent connection.

Because of the persistent connection, dedicated streaming protocols allow random access (\eg fast forward) within the streamed video. In addition, they allow adaptive streaming, in which multiple encoded video streams could be delivered to different players based upon available bandwidth and processing characteristics. The streaming server can monitor the outbound flow, so if the viewer stops watching the video, it stops sending video packet to the viewer. 

Video content in the streaming server is split  into small chunks, whenever these chunks are sent to the viewers, they are cached at the local device and will be removed after they are played. This feature offers freedom to viewer to move back and forth within the streamed video. It also protects the copyright of the video content. 
Although streaming technology was attractive in the beginning, its drawback appeared after deployment. A streaming protocol, \eg RTMP~\cite{lesser2005programming} used by Adobe Flash, utilizes different port numbers from HTTP. As such, RTMP packets can be blocked with some firewalls. The persistent connection between the streaming server and viewer players increases the network usage and causes limited scalability for streaming servers. 


\subsubsection{Adaptive Streaming} 
To address the limitations of previous streaming technologies, HTTP-based streaming solutions came back to the forefront of streaming technology-adaptive streaming. All adaptive streams are broken into chunks and stream separate videos. There is no persistent connection between the server and the player. Instead of retrieving a single large video file in one request, adaptive streaming technology retrieves a sequence of short video files in an on-demand basis.

Adaptive streaming has the following benefits: First, like streaming server, there is no network wastage, because the video content is delivered on the go. Therefore, one HTTP server can efficiently serves several streams. Second, HTTP-based streaming is delivered through HTTP protocol, which avoids the firewall issue faced by RTMP. Third, it costs less than using streaming server. Fourth, it can scale quickly and effectively to serve more viewers. Fifth, seeking inside the stream is no more an issue. When the viewer moves the player forwarder, the player just retrieves the exact video segments as opposed to the entire video up to the requested point. 

There are four main adaptive streaming protocols, namely MPEG Dynamic Adaptive Streaming over HTTP (DASH)~\cite{stockhammer2011dynamic}, Apple HTTP Live Streaming (HLS)~\cite{thang2014evaluation}, and Microsoft Smooth Streaming~\cite{begen2011watching}. 

MPEG DASH delivers ISO Base Media File Format (ISOBMFF)~\cite{bouzakaria2014overhead} video segments. It defines a Media Presentation Description (MPD) XML document to provide the locations of video streams, so that players know where to download them. The media segments for DASH is delivered with formats either based on the ISOBMFF~\cite{timmerer2010http} or Common Media Application Format (CMAF) standards~\cite{cmaf}. 

Apple's HLS is well-known and implemented on the Apple devices (and nowadays on Android devices too). It utilizes a M3U8 master manifest to include multiple media playlists, each playlist represents one stream version and it contains the location of all the video segments for this stream. HLS video segment uses either MPEG-2 Transport Stream (M2TS)~\cite{hanna1995demultiplexer} or CMAF~\cite{cmaf} for H.264 encoded videos, and ISOBMFF for HEVC encoded videos.

Smooth Streaming has two separated manifest files, namely SMIL server manifest file and client manifest file. They are all defined in XML format documents. Smooth streaming also delivers video segment with a format (known as ISMV) based on ISOBMFF.


These four adaptive streaming technologies have empowered streaming service providers to deliver the video contents to viewers smoothly even under low bandwidth Internet connection. However, to support all viewers' platforms, stream providers have to deploy and maintain all these four streaming protocols that subsequently increases complexity and costs. The supported platforms of these four protocols are shown in Table~\ref{tbl:platform}

\renewcommand{\arraystretch}{1.5}
\begin{table}[ht]
  \centering
  \caption{\label{tbl:platform}Adaptive Streaming Supported Platforms}
  \resizebox{\columnwidth}{!}{
    \begin{tabular}[t]{lccc}
      \toprule
        & \textbf{Desktop Player} & \textbf{Mobile Device Support} & \textbf{OTT Support} \\ 
      \midrule
        MPEG-DASH & dash.js, dash.as, GPAC & Windows, Android Phone & Google TV, Roku, Xbox 360 \\ 
        Apple HTTP Live Streaming (HLS) & iOS, Mac OSX, Flash & iOS/Android3.0+ & Apple TV, Boxee, Google TV, Roku \\ 
        Smooth Streaming & Silverlight & Windows Phone & Google TV, Roku, Xbox 360 \\
      \bottomrule
    \end{tabular}
  }
\end{table}

\subsection{Video Streaming Delivery Networks}\label{sec:delivery_content}
\subsubsection{Content Delivery Networks (CDN)}
The goal of CDN technology is to reduce the network latency of accessing web contents. CDNs replicate the contents to geographically distributed servers that are close to viewers~\cite{saro,vakali}. Considering that large size of video contents usually takes a long transmission time, using CDNs to cache video contents close to viewers reduces the latency  dramatically. Netflix, as one of the largest stream providers use three different CDN providers (Akamai, LimeLight, and Level-3) to cover all viewers in different regions~\cite{adhikari}. The transcoded and packaged video contents are replicated in all three CDNs. 

Streaming video contents through CDN has been studied in earlier works \cite{john, cranor, wee}. Cranor \etal~\cite{cranor} proposed an architecture (called PRISM) for distributing, storing, and delivering high-quality streaming content over IP networks. They proposed a framework to support VOD streaming content via distribution networks services (\ie CDNs). Wee \etal~\cite{wee} presented an architecture for mobile streaming CDN which was designed to fit the mobility and scaling requirements. Apostolopoulos \etal~\cite{john} proposed multiple paths between nearby edge servers and clients in order to reduce latency and deliver high-quality streaming. Benkacem \etal~\cite{benkacem2018performance} provided an architecture to offer CDN slices through multiple administrative cloud domains. Al-Abbasi \etal~\cite{al2018edgecache} proposed a model for video streaming systems, typically composed of a centralized origin server, several CDN sites, and edge-caches located closer to the end users. Their proposed approach focused on minimizing a performance metric, stall duration tail probability (SDTP), and present a novel and efficient algorithm accounting for the multiple design flexibilities. The authors demonstrated that the proposed algorithms can significantly improve the SDTP metric, compared to the baseline strategies.

Live streaming CDN is discussed in~\cite{kontothanassis,yin, lu2011scalable} to improve scalability, latency quality, and reliability of the service. 



\subsubsection{Peer to Peer (P2P) Networks}
 P2P networks enable direct sharing of computing resources (\eg CPU cycles, storage, and content) among peer nodes in a network~\cite{ramzan2012video}.
P2P networks are designed for both clients and servers to act as peers. They can download data from same nodes and upload them to other nodes in the network. P2P networks can disseminate data files among users within a short period of time. P2P networks are extensively utilized for video streaming services as well.

 P2P streaming is categorized into two types, namely tree-based and mesh-based. Tree-based P2P structure distributes video streams by sending data from a peer to its children peers. In mesh-based P2P structure, the peers do not follow a specific topology, instead, they function based on the content and bandwidth availability on peers~\cite{liu2008survey}. One drawback of using tree-based P2P streaming is the vulnerability to peers' churn. The drawback of deploying mesh-based P2P streaming structure is video playback quality degradation, ranging from low video bit rates and long startup delays, to frequent playback freezes. Golchi \etal~\cite{golchi2018evaluation} proposed a method for streaming video in P2P networks. Their approach uses the algorithm of improved particle swarm optimization(IPSO) to determine the appropriate way for transmitting data that ensures service quality parameters. They showed that the proposed approach works efficiently more than the other related methods.

\begin{figure}
        \begin{subfigure}[b]{0.5\textwidth}
                \centering
                \includegraphics[width=.85\linewidth]{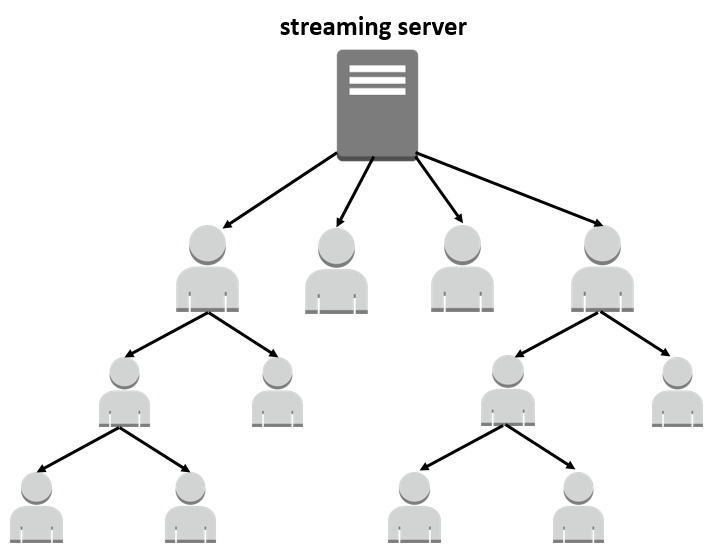}
                \caption{single tree-based streaming}
                \label{fig:single-tree}
        \end{subfigure}%
        \begin{subfigure}[b]{0.5\textwidth}
                \centering
                \includegraphics[width=.85\linewidth]{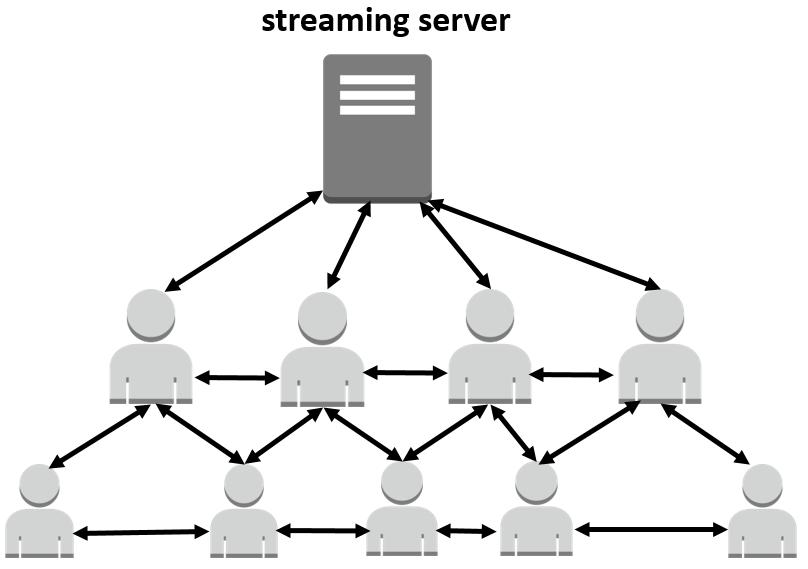}
                \caption{ mutli-tree streaming}
                \label{fig:mesh-tree}
        \end{subfigure}%
\caption{Ways to achieve peer-to-peer (P2P) video streaming}
\label{fig:p2p-tree}
\end{figure}

 Figure \ref{fig:p2p-tree} shows two popular variations of tree-based P2P structure, namely single tree-based and multi-tree based structures. In \emph{Single Tree-based streaming}, shown in Figure \ref{fig:single-tree}, video streaming is carried out at the application layer and the streaming tree is formed by participating users. Each viewer joins the tree at a certain level and receives the video content from its parent peer at the above level. Then, it forwards the received video to its children peers at the below level~\cite{chu2000case, jannotti2000overcast}.

In \emph{Multi-Tree Streaming} (aka mesh-tree), shown in Figure \ref{fig:mesh-tree},  a peer  develops the peering connection with other close peers. Then, the peer can potentially download/upload video contents from/to multiple close peers simultaneously. If a peer loses the connection to the network, the receiving peer could still download video from remaining close peers. Meanwhile, the peer can locate new peers to maintain a required level of connectivity. The strong peering in mesh-based streaming systems supports them to be robust against peer churns~\cite{magharei2007mesh, venkataraman2006chunkyspread}. Ahmed \etal~\cite{ahmad2018peer}proposed a multi-level multi-overlay hybrid peer-to-peer live video system that offers to the Online Games players the way of streaming the video simultaneously and enables to watch the game videos of other players. Their approach aimed to reduce the transmission rate while increasing the number of peers and the delivery and reliability of data are guaranteed on time.

Tree-based P2P VOD streaming shares the video stream on the network and achieves fast streaming. The video stream is divided into small size data block. The server disperses the data blocks to different nodes. The nodes download their missing blocks from their neighboring peers~\cite{vlavianos2006bitos}

Guo \etal~\cite{guo2003p2cast} proposed an architecture that uses the P2P approach to cooperatively stream video by only relying on unicast connections among peers. Xu \etal~\cite{xu2008balanced} proposed an approach based on binary tree strategy to distribute VOD P2P networks. Their approach divides the videos into segments to be fetched from several peers. Yiu \etal~\cite{yiu2007vmesh} proposed VMesh distributed P2P VOD streaming approach. In their approach, videos are split into segments and then stored at the storage of the peers locally. Their proposed design presents to peers an ability to forward, backward, pause, and restart during the playback. Cheng \etal~\cite{cheng2007supporting} proposed a topology enhancing video cassette recording (VCR) functions for VOD services in the networks.  Their approach allows  a peer to achieve fast  seeks relocation by keeping close neighbors and remote them in a set of rings. Xu \etal~\cite{xu2009supporting} proposed a scheme based tree to make user interacting in the P2P streaming. The proposed scheme presents an advantage to support the users requests asynchronously while maintaining high resilience. 

The main advantages of P2P are low cost and flexibility for scalability, however, it suffers from instability in QoS. Therefore, researchers proposed to combine the advantages of both CDN and P2P in one system. Accordingly, several hybrid systems were developed to combine P2P and CDN for content streaming. Afergan \etal \cite{afergan2012hybrid} proposed an approach which utilizes CDN to build CDN-P2P streaming. In their proposed design, they optimized dynamically the number and locations of replicas for P2P service. Xu \etal \cite{xu2006analysis} presented a scheme that is formed by both CDN and P2P streaming. They showed the efficiency of their approach by reducing the cost of CDN without impacting the quality of the delivered videos.

\subsection{Video Streaming Security}\label{sec:security}
\subsubsection{Privacy}
With the ubiquity of video streaming on a wide range of display devices, the privacy of the video contents has become a major concern. In particular, live contents either in form of video surveillance or user-based live-streaming capture places and record many unwanted/unrelated contents. For instance, a person who live-streams from a street, unintentionally may capture plate number of vehicles passing that location. Therefore, video streaming systems can compromise the privacy of people (\eg faces and vehicles tags). Various techniques have been developed to protect the privacy of live video contents. 

Dufeaux \etal \cite{dufaux2008scrambling} introduced two techniques to obscure the regions of interests while video surveillance systems are running. 
 Zhang \etal \cite{zhang2005hiding} came up with a framework to store the privacy information of a video surveillance system in form of a watermark. The proposed model embeds a signature into the header of the video. Moreover, it embeds the authorized personal information into the video that can be retrieved only with a secret key. Another research, conducted by Carrillo \etal \cite{carrillo2008compression}, introduces a compression algorithm for video surveillance system, which is based on the encryption concept. The proposed algorithm protects privacy by hiding the identity revealing features of objects and human. Such objects and human identity could be decrypted with decryption keys when an investigation is requested.
 
Live video contents commonly are transmitted via wireless media which can be easily intercepted and altered. Alternatively, DOS attacks can be launched on the live video traffics~\cite{hurani17}. As such, in~\cite{Alsmirat2017,Feher06} algorithms are provided to distinguish between packets that were damaged because of noise or an attack. In the former case, the errors must be fixed while in the latter the packet must be resent. The algorithm counts the number of 1s or 0s in a packet before the packet is send and uses that count to generate a message authentication code (MAC). The MAC is appended to the end of the packet and sent over the network. When the packet is received, the MAC
is calculated again, and the two codes are compared. If the differences in MACs is past a certain threshold, the past is marked as malicious and discarded. Because the MAC is also sent over the network, the algorithm will detect bit errors in both the packet and the MAC.

\subsubsection{Digital Rights Management}
Another security aspect in video streaming is the copyright issue. This is particularly prominent for subscription-based video streaming services (such as Netflix).
Digital Rights Management (DRM) is the practice of securing digital contents, including video streams, to prevent unlawful copying and distribution of the protected material. Specifically, its goals are typically centered around controlling access to the content, enforcing rules about usage, and identifying and authenticating the content source.  As such, contents protected with DRM are typically sold as a license to use the contents, rather than the content itself.  DRM solutions to meet these goals often incorporate tools such as encryption, watermarking, and cryptographic hashing \cite{drm:watermark:multiparty}.

\begin{figure}
\includegraphics[width=4in]{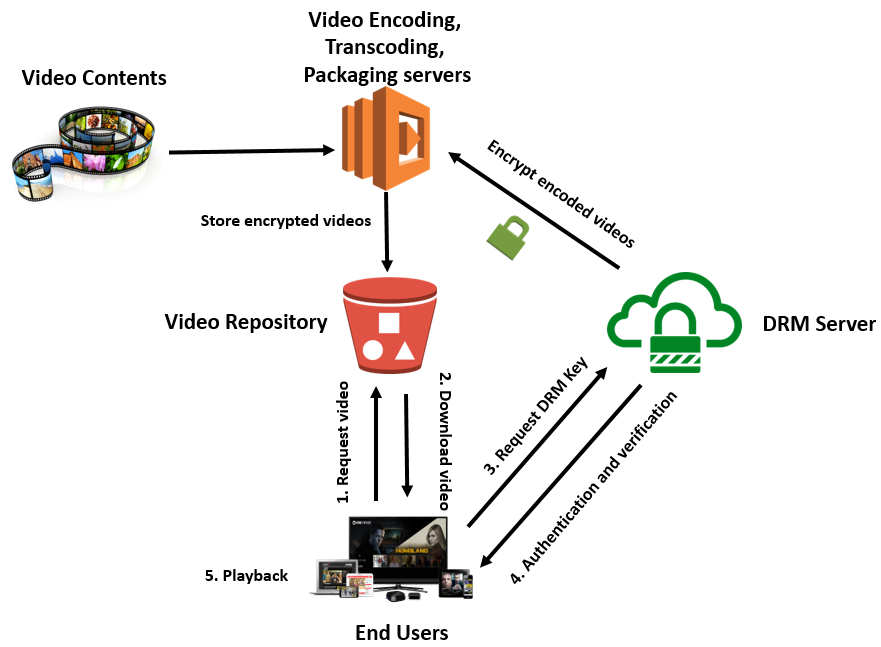}
\caption{Workflow of Digital Right Management (DRM) to support of video streaming security}
\label{fig:drm}

\end{figure}

The process of DRM starts with encryption, video contents are encrypted and stored in the video repository and DRM providers keep the secret keys. Fig.~\ref{fig:drm} summarizes the steps to stream a video protected with DRM. Upon request to stream an encrypted video, after downloading it, the secret key is requested from the DRM provider server to decrypt the video.

Currently, there are three major DRM technologies, namely Apple's Faireplay~\cite{fairplay},
Microsoft's PlayReady~\cite{kumar2013drm} and Google's Widevine~\cite{kumar2013drm}. These technologies are able to serve most of the devices and platforms in the market. To cover all devices, streaming service providers have to encrypt every video content with three different DRM technologies and store them in the repository. The three DRM technologies supported platforms are shown in Table~\ref{tbl:drm}.

\renewcommand{\arraystretch}{1.5}
\begin{table}[ht]
  \centering
  \caption{\label{tbl:drm}DRM Supported Platforms}
  \resizebox{\columnwidth}{!}{
    \begin{tabular}[t]{lcccccccccccc}
      \toprule
        & Chrome & FireFox & IE 11 & Safari & Android & iOS & Windows Phone & ChromeCast & Roku & Apple TV & Xbox  \\
      \midrule
         Fairplay &  & &  & \ding{51} & & \ding{51} &  &  &  & \ding{51} &  \\ 
         PlayReady  &  &  & \ding{51} & \ding{51} &  &  & \ding{51} & \ding{51} & \ding{51} & & \ding{51} \\ 
         Widevine & \ding{51} & \ding{51} & &  &  \ding{51} &  &  & \ding{51} &  &  &  \\ 
      \bottomrule
    \end{tabular}
  }
\end{table}

With the increasing demand to stream DRM-protected VOD, an offline application-driven DRM interceptor has been proposed in~\cite{dorwin2015application} that is the awareness of the network connection status of the client device. In this case, if the interceptor application decides that the client device is offline,  it requests the license/key for the protected content. The license/key is controlled by the interceptor application. Accordingly, the requests of license/key are handled by the interceptor application that retrieves them from a locally data-store, and  then send the key/license to the DRM module.

\subsection{Analysis of Video Streaming Statistics}\label{sec:analysis}
\subsubsection{Impact of quality on viewers' behavior}
Previous studies show that the quality of video streaming impacts on viewers' reception to the video and the revenue of the stream provider~\cite{krishnan2013video}. According to the studies, starting video streaming with delay and interruption during video streaming significantly increases the possibility of abandoning watching the videos.

Florin \etal ~\cite{dobrian2011} addressed the impact of video quality on viewers' interest. They claimed that the percentage of time spent on buffering (\ie buffering ratio) has the largest impact on the user interest across all content types. The average bit-rate of live streaming has a significant impact on user abandonment of watching the video.
Video streaming providers should attempt to maximize viewer engagement by minimizing the buffering time and rate and increasing the average bit-rate. 




\subsubsection{Access Pattern to Video Streams}
Analysis of access pattern to videos shows that the access pattern of videos streams in a repository follows a long-tail distribution~\cite{intro_9}. That is, few popular videos, known as \emph{hot} videos that construct around 5\% of the repository, are accessed very often while a large portion of non-popular videos are accessed infrequently.
  
The studies also reveal that viewers are typically interested in recently posted videos. 
Moreover, for new videos, the popularity fluctuates significantly while the popularity of old videos does not fluctuate significantly~\cite{cha2007}.
  
Video access rate indicates how many times a video is accessed by a user, however, it does not implicate if the accessed video stream is played to end of it or not. In fact, recent studies (\eg \cite{miranda2013characterizing}) showed that, the beginning segments of a video stream are played more frequently than the rest of it. Miranda et al. \cite{miranda2013characterizing} revealed that in a video stream, the views are distributed  following a long tail distribution. More specifically, the distribution of views of the segments (GOPs) in a video stream can be calculated by the Power-law \cite{newman2005power} model.
  
The access rate of GOPs in all video streams in a repository does not necessarily  follow long-tail pattern as stated earlier. There are video streams whose some GOPs in the middle or end of the video stream are accessed more frequently than other GOPs. An example of a soccer match streaming can show GOPs with a tremendously higher access rate where a player scores a goal. We define this type of video streams as those with non-long-tail access pattern \cite{darwich2017}. 

\subsubsection{Video Streaming based Recommendation}
As mentioned above, provided the access pattern of video streams for a given viewer, stream providers are able to predict the video categories the viewer prefers and recommend them at the front page. The same strategy can be used for video recommendation to the viewers located in the same geographical area. To improve the accuracy of prediction and recommendation, machine learning~\cite{nasrabadi2007pattern} and deep learning~\cite{lecun2015deep} approaches have widely been applied for this purpose. The most successful example is Netflix and YouTube machine-learning-based recommendation systems as explained in~\cite{gomez2016netflix}.

\subsection{Storage of Video Repositories}\label{sec:storage}
Video streaming repositories are growing in size, due to the increasing number of content creation sources, diversity of display devices, and the high qualities viewers desire. This rapid increase in the size of repositories implies several challenges for multimedia storage systems. In particular, video streaming storage challenges are threefold: capacity, throughput, and fault tolerance.  

One of the main reasons to have a challenge in video streaming storage capacity is the diversity of viewers' devices. To cover increasingly diverse display devices, multiple (more than 90) versions of a single video should be generated and stored.However, storing several formats of the same video implies a large-scale storage system. 
Previous studies provide techniques to overcome storage issues of video streaming. Miao \etal~\cite{miao2002} proposed techniques to store some frames of a video on the proxy cache. Their proposed technique reduces the network bandwidth costs and increases the robustness of streaming video on poor network conditions.

When a video is stored on a disk, concurrent accesses to that video are limited by the throughput of the disk. This restricts the number of simultaneous viewers for a video. To address the throughput issue, several research studies have been undertaken. Provided that storing video streams on a single disk has a low throughput, multiple disk storage are configured to increase the throughput. Shenoy \etal~\cite{shenoy} propose data stripping where a video is segmented and saved across multiple storage disks to increase the storage throughput.
Videos streams are segmented into blocks before they are stored. The blocks can be stored one after another (\ie contiguously) or scattered over several storage disks. Although contiguous storage method is simple to implement, it suffers from the fragmentation problem. The scattered method, however, eliminates the fragmentation problem with the cost of more complicated implementation. 

Scattering video streams across multiple disks and implementing data striping and data interleaving methods improves reliability and fault tolerance of video streaming storage systems~\cite{wu}.

\section{Cloud-based Video Streaming}\label{sec:cbvs}
\subsection{Overview}

In this section, we discuss how offered cloud services can be useful for different processes in video streaming. Broadly speaking, we discuss computational services, networking services, and storage services offered by cloud providers and are employed by video stream providers. We also elaborate on possible options within each one of the cloud services.

\subsection{Computing Platforms for Cloud-based Video Streaming}

Upon arrival of a streaming request, the requested video stream is fetched from the cloud storage servers and the workflow of actions explained in Section~\ref{subsec:workflo} are performed on them, before streaming them to the viewers. These processes are commonly implemented in form of independent services, known as micro-services~\cite{newman2015building}, and are generally deployed in a loosely coupled manner on independent servers in cloud datacenters. Services like web server, video ingestion, encoding, transcoding, and packaging are examples of micro-services deployed in datacenters for video streaming.
For the sake of reliability and fault tolerance, each of these micro-services is deployed on multiple servers and form a large distributed system. A load balancer is deployed for the servers of each Microservice. The load balancer assigns streaming tasks to an appropriate server with the goal of minimizing latency and to cover possible faults in the servers. 


The aforementioned micro-services are commonly implemented via container technologies~\cite{thones2015microservices}, possibly using serverless computing paradigm~\cite{serverless18}. Docker containers scale up and down much faster than VMs and have faster startup (boot up) times that gives them an advantage to VMs in handling fluctuating video streaming demands. In addition, Docker containers are used in video packaging, handling arriving streaming requests (known as request ingestion), and inserting advertisements within/between video streams. 

Sprocket~\cite{serverless18} is a serverless system implemented based on AWS Lambda~\cite{usenixserverless18} and enables developers to program a series of operations over video content in a modular and extensible manner. Programmers implement custom operations, ranging from simple video transformations to more complex computer vision tasks, and construct custom video processing pipelines using a pipeline specification language. Sprocket then manages the underlying access, encoding and decoding, and processing of video and image content across operations in a parallel manner. Another advantage of deploying serverless (aka function-as-a-service) computing paradigm, such as AWS Lambda, for video streaming is to relieve stream providers from scheduling, load balancing, resource provisioning, and resource monitoring challenges.

Apart from the type of computing platform (\eg VM-based and container-based) for video streaming, the type of machines provisioned to process streaming tasks are also influential in the latency and incurred cost of streaming. For instance, clouds offer various VM types with diverse configurations (\eg GPU base, CPU base, IO base, and Memory base), or various reservation types (\eg on-demand, spot, and advance reservation).

\subsection{Cloud-based Video Transcoding}
To cover viewers with heterogeneous display devices that work with diverse codecs, resolutions, bit-rates, and frame rates, video contents usually have to be transcoded to several formats.
Video transcoding process takes major computing power and time in the whole video streaming workflow~\cite{tpds17}. The Methods and challenges for video transcoding have been studied by Vetro \etal~\cite{intro_7} and Ahmad \etal~\cite{intro_6}. In the past, streaming service providers had to maintain large computing systems (\ie in-house datacenters) to achieve video transcoding. However, due to the update and maintenance costs of in-house datacenters, many streaming service providers have chosen to outsource their transcoding operations to cloud servers~\cite{tpds17,li2018cost,pre_3,pre_5,mahmoud17}. Extensive computational demand of video transcoding can potentially impose a significant cost to stream providers. As such, it is important that stream providers apply proactive methods to transcode videos in their repositories.

\begin{figure}[htb] 
    \centering
    \includegraphics[width=5.5in]{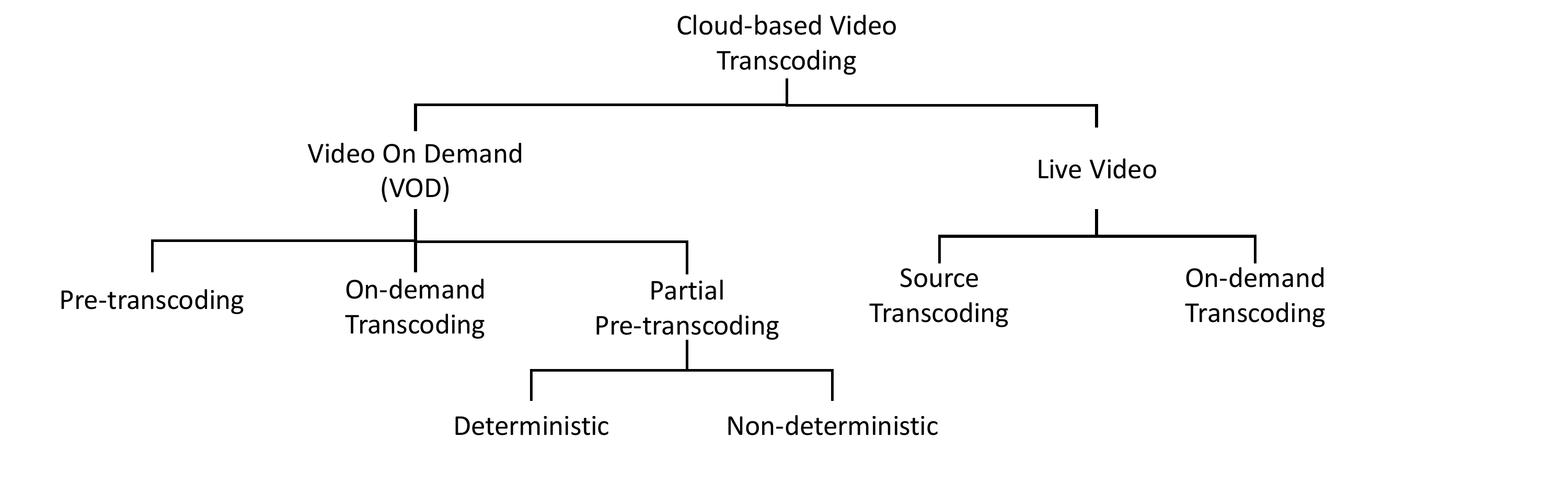}\vspace{-3mm}
    \caption{Approaches to perform video transcoding using cloud services.}
    \label{fig:rw}
\end{figure}

A taxonomy of various cloud-based video transcoding is shown in Figure~\ref{fig:rw}. Cloud-based solutions for transcoding VODs are either based on creating several versions of the original video in advance (aka pre-transcoding~\cite{rw_12, bg_2, rw_3, rw_10, rw_11}) or transcoding videos upon viewer's request (aka on-demand transcoding~\cite{pre_3, pre_4, li2012cloud}). In considering the long-tail access pattern to video streams on the Internet has made the on-demand approach an attractive option for stream providers. However, pure on-demand transcoding approach can increase latency and even the incurred cost for stream providers~\cite{mahmoud17}. Therefore, approaches have been introduced to perform pre-transcoding in a more granular level. That is, pre-transcoding only parts of a video streams (\eg few important GOPs). This approach is known as partial pre-transcoding~\cite{darwich2017}. 

According to Darwich \etal~\cite{darwich2017}, partial pre-transcoding can be carried out either in a deterministic or non-deterministic manner. Considering the fact that the beginning of video streams are generally watched more often~\cite{miranda2013characterizing}, in the deterministic approach, a number of GOPs from the beginning of the video are pre-transcoded, and transcode the rest in an on-demand manner~\cite{mahmoud17}. Alternatively, in the non-deterministic approach~\cite{darwich2017}, pre-transcoding is not limited to the beginning of video streams and can be performed on any popular GOP, disregarding its position in the stream. Although the non-deterministic approach is proven to be more efficient, it imposes the overhead of maintaining and processing view metadata for each GOP.

Procuring various video formats in live streaming can be achieved by camera, upon video production (known as source transcoding). However, due to the high latency and inefficiency of source transcoding, live videos are commonly transcoded in the run time~\cite{pre_5, rw_22, thang2012adaptive}. 

Research have been focused on video segmentation~\cite{rw_12, bg_2}, load balancing~\cite{rw_3, rw_11}, and resource provisioning~\cite{rw_3, rw_10}, with the goal of maximizing throughput of the pre-transcoding operation. Alternatively, in on-demand transcoding, the goal is to respect QoE of viewers in form of minimizing the response time of the transcoding operation. More specifically, each segment (GOP) of the video needs to be transcoded before its presentation time (aka deadline).

To use cloud services efficiently, it is crucial to realize the performance difference among video transcoding operations on different VM types. Li \etal~\cite{tpds17} have evaluated the performance of various transcoding operations on heterogeneous VMs to investigate the key factors on transcoding processing time. They identified the affinity (\ie match) of various VM types (\eg GPU versus CPU and Memory-based) with different transcoding operations. They show that GOP frame numbers is the most influential factor on transcoding time. In fact, frame numbers in a GOP can imply its content type. Videos with fast-motion content include numerous GOPs with few frames and vice versa. The execution time of each small GOP is usually low, and therefore it can be transcoded on cost-efficient VM types without compromising performance. 
In addition, Li \etal~ provide a suitability model of heterogeneous VM types for various transcoding operations in consideration of both performance and cost factors.  Transcoding time estimation is needed to improve the quality of scheduling and VM provisioning on the cloud. Deneke \etal~\cite{deneke2014video} propose a machine learning model to predict the transcoding time based on the video characteristics (\eg resolution, frame rate, and bit rate). 

To further cut the cost of utilizing cloud services for on-demand video transcoding, Li \etal~\cite{pre_3,li2018cost} propose Cloud-based Video Streaming Engine (CVSE) that operates at the video repository level and pre-transcodes only hot videos while re-transcodes rarely accessed videos upon request. Li \etal define desired QoE of streaming as minimizing video streaming startup delay via prioritizing the beginning part of each video. The engine is able to provide low startup delay and playback jitter with proper scheduling and resource provisioning policy. 

Transcoding video in an on-demand way does reduce the storage cost, however, it incurs computing cost. How to balance these two operations to minimize the incurred cost is another challenge. Jokhio \etal~\cite{rw_13} presents a trade-off method to balance the computation and storage cost for cloud-based video transcoding. The method is mainly based on the time and frequency for a given video to be stored and re-transcoded, respectively. Compared to Jokhio's work, Zhao~\etal~\cite{rw_18} also take the popularity of the video into consideration. Darwich \etal~\cite{darwich2017} propose a method to partially pre-transcode video streams depending on their degree of hotness. For that purpose, they define a method based on the hotness of each GOP within the video. 

Barais~\etal~\cite{barais2016towards} propose a Microservice architecture to transcode videos at a lower cost. They treat each module (\eg splitting, scheduling, transcoding, and merging) of transcoding as a separate service, and running them on different dockers. To reduce the computational cost of on-demand transcoding, Denninnart \etal~\cite{chavit18} propose to aggregate identical or similar streaming micro-services. Identical streaming micro-services appear when two or more viewers stream the same video with the same configurations. Alternatively, similar streaming micro-services appear when viewers stream the same video but with different configurations or even different operations. An example of identical micro-services can be when two or more viewers stream the same video for the same type of device (\eg smart-phone). However, when the viewers stream the same video on distinct devices (\eg smart-phone versus TV) that have different resolution characteristics, the video has to be processed to create two different resolutions, which creates micro-services with various configurations. Interestingly, during peak times the method becomes more efficient because it is more likely to find similarity between micro-services.


Cloud-based video transcoding has also been widely used in live streaming~\cite{pre_5, rw_22}. Lai \etal~\cite{rw_22} propose a cloud-assisted real-time transcoding mechanism based on HLS protocol, it can analyze the online quality between client and server without changing the HLS server architecture, and provides the good media quality. Timmerer \etal~\cite{timmerer2015live} present a live transcoding and streaming-as-a-service architecture by utilizing cloud infrastructure, it is able to take the live video streams as input and output multiple stream versions based on the MPEG-DASH~\cite{thang2012adaptive} standard.



\subsection{Cloud-based Video Packaging}

Video packaging is computationally lighter than video transcoding, therefore, it is beneficial and more feasible to process them in an on-the-fly by utilizing cloud services (VMs or containers)~\cite{myers2017methods}. For VOD streaming, video contents are usually pre-transcoded to different renditions (formats), then each rendition is packaged into various versions to meet different streaming protocol requirements (\eg DASH, HLS, Smooth). 

Instead of statically packaging transcoded videos into different protocol renditions and store them in the repository (\ie pre-packaging), dynamic video packaging packages video segments based on the device's supported protocols. The whole process only takes milliseconds~\cite{gorostegui2017broadcast}, and viewers do not generally notice it, however, it saves a significant storage cost for video streaming providers.

Due to the lightweight nature of the packaging process, container services are generally used for their implementation in cloud data centers.

\subsection{Video Streaming Latency and Cloud-based Delivery Networks}


Clouds are known for their centrality and high latency communication to users~\cite{razin18}. To reduce the network latency and to reduce the load of requests from the servers, video contents are normally cached in the Content Delivery Networks (CDN) such as Akamai. The CDNs are distributed and located geographically close to viewers. 

\begin{figure}[htb] 
    \centering
    \includegraphics[width=5in]{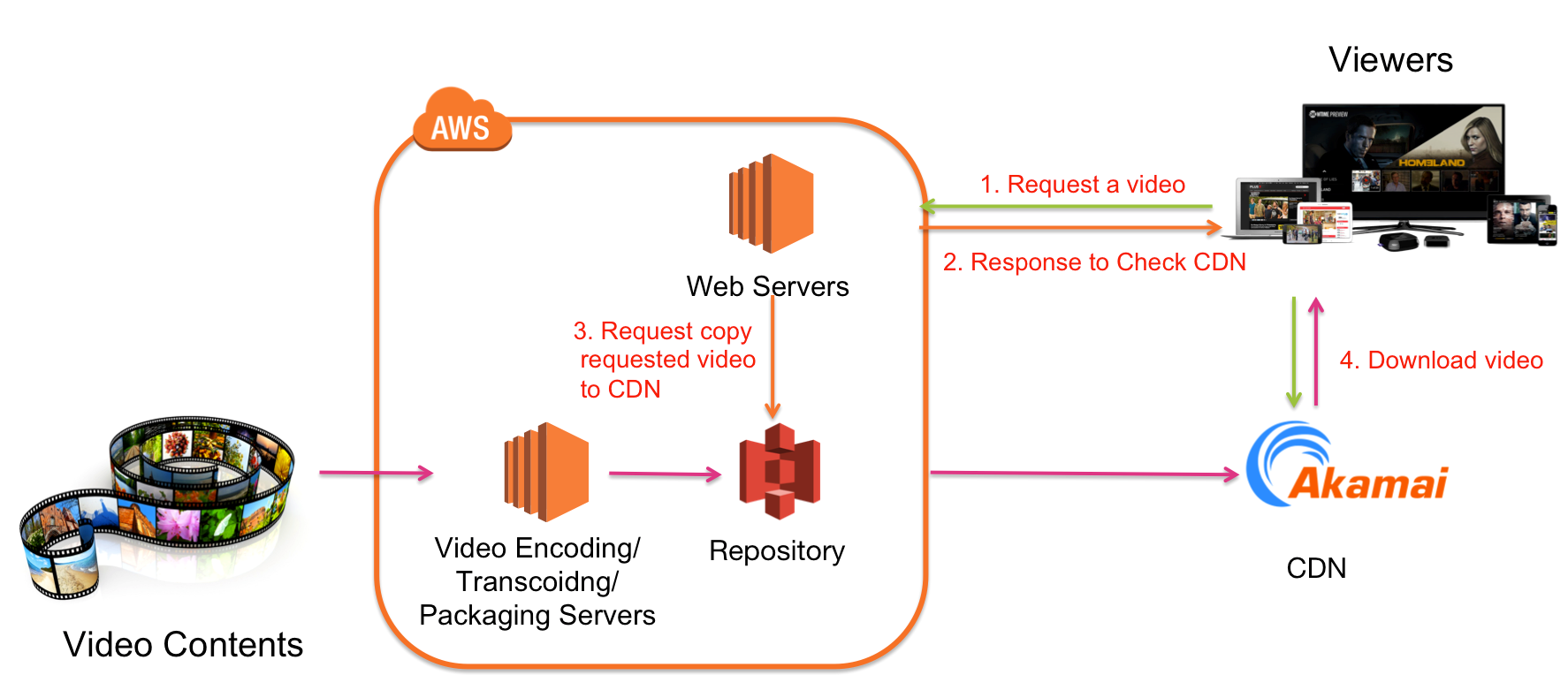}
    \caption{Workflow of actions taken place for low latency video streaming}
    \label{fig:cs}
\end{figure}

A workflow of the whole video streaming process is shown in Figure~\ref{fig:cs}. Upon a viewer's request to stream a video, the request is first received by a web server in the cloud datacenter. If the requested video is already cached in the CDN, the web server will send back a manifest file which informs the viewer's computer about the CDN that holds the video files. Then, the viewer sends another request to the CDN and stream the video. However, if the requested video is not cached in the CDN, the web server has to process and send the content from cloud storage to the CDN. Then, the server sends a copy of the file that includes the CDN address to the viewer to start streaming from the CDN.

Content delivery networks (CDNs) that are offered in form of a cloud service are known as cloud CDN. Compared to traditional CDNs, cloud CDNs are cost-effective and offer low latency services to content providers without having their own infrastructure. The users are generally charged based on the bandwidth consumption and storage cost~\cite{chen2012intra}. 

Hu \etal~\cite{hu2016joint} presented an approach using the cloud CDN to minimize the cost of system operation while maintaining the service latency. Based on viewing behavior prediction, Hu \etal~\cite{hu2014community} investigated the community-driven video distribution problem under cloud CDN and proposed a dynamic algorithm to trade-off between the incurred monetary cost and QoS. Their results came with less operational cost while satisfying the QoS requirement.

Jin \etal~\cite{jin2012codaas} proposed a scheme that offers the service of on-demand virtual content delivery for user-generated content providers to deliver their contents to viewers. The proposed approach was developed using a hybrid  cloud. Their scheme offered  elasticity and privacy by using virtual content delivery service with keeping the QoE to user-generated content providers.

Li \etal~\cite{li2011cost} proposed  an approach for partial migration of VoD using the hybrid cloud deployment. Their proposed solution allows the requests of user to be partly served based on the self-owned servers and partly used the cloud.  their Proposed migration approach (active, reactive, and smart strategies) helps the hybrid cloud to save up to 30\% bandwidth expenses compared to the client/server mode. Researchers at Microsoft conducted an experiment within Microsoft public cloud CDN, Windows Azure, to demonstrate the benefits of CDN integration with the cloud. The results show a significant gain in large data download by utilizing CDN in Cloud Computing \cite{li2012utilizing}.

\subsection{Cloud Storage for Video Streaming}

\renewcommand{\arraystretch}{1.5}
\begin{table}[ht]
  \centering
  \caption{\label{ref:storing-technologies}Comparison of different technologies for storing video stream}
  \resizebox{\columnwidth}{!}{
    \begin{tabular}[t]{lcccccc}
      \toprule
        &\textbf{Accessibility}  &\textbf{Capacity}  & \textbf{Scalability} & \textbf{Reliability} & \textbf{Cost} & \textbf{QoE}\\
      \midrule
         In House Storage & limited & large & no & no & high & high \\
         Cloud Storage (Outsourcing) &high  &large&yes&yes& low&low  \\ 
         Content Delivery Network (CDN) &high  &small&no&yes&high&high   \\ 
         Peer 2 Peer &high&large&yes&yes&low&low \\
         Hybrid CDN-P2P & high&small& yes&yes&low&high\\
      \bottomrule
    \end{tabular}
  }
\end{table}

The rapid growth of video streaming usage in various applications, such as e-learning, video surveillance and situational awareness, and on various forms of mobile devices (\eg smart-phones, tablets, laptops) has created the problem of \emph{big video data}~\cite{bigvideo17}. The fast growth of video contents on the Internet requires massive storage facilities. However, the current storage servers face scalability and reliability issues, in addition to the high maintenance and administration cost for storage hardware.
The cloud storage services provide a solution for scalability, reliability, and fault tolerance~\cite{rodriguez}. As such, major streaming service providers (\eg Netflix and Hulu) have relied entirely on cloud storage services for their video storage demands.

Table~\ref{ref:storing-technologies} provides a comparison of different storage solutions for video streaming in terms of accessibility of viewers to the same video stream, available capacity  (storage space), scalability, reliability, incurred cost, and QoS. It is worth noting that although CDN is not a storage solution, but it can be used to reduce the storage cost by caching temporally hot video streams. Therefore, we consider it in our comparison table.

On-demand processing of video streaming is one effective method to reduce the storage volume and cost. This is particularly important when the long-tail access pattern to video streams is considered~\cite{darwich2017}. That is, except for a small portion of video streams that are hot the rest of videos are rarely accessed. Gao \etal~\cite{gao} propose an approach that partially pre-transcodes  video contents in the Cloud. Their approach pre-transcodes the beginning segments of video stream and  which are more frequently accessed, while transcoding the remaining contents video stream upon request, this results to a reduction of storage cost.
They demonstrated that their approach reduces 30\% of the cost compared to pre-transcoding the whole contents of video stream.


Darwich \etal~\cite{darwich2017} proposed a storage and cost-efficient method for cloud-based video streaming repositories based on the long-tail access patterns. They proposed both repository and video level solutions. In the video level, they consider access patterns in two cases, (A) when it follows a long-tail distribution; and (B) when the video has random (\ie non-long-tail) access pattern. They define the cost-benefit of pre-transcoding for each GOP and determine the GOPs that need to be pre-processed and the ones that should be processed in an on-demand manner.
 
Krishnappa \etal~\cite{krishnappa} proposed strategies to transcode  the segments of a video stream requested by users. To keep  a minimized startup delay of video streams when applying online strategies on video, they came up with an approach to predict the next video segment that is requested by the user. They carried out their prediction approach by implementing Markov theory. Their proposed strategy results a significant reduction in the cost of using the cloud storage with high accuracy.


\section{Summary and Future Research Directions}\label{sec:conclusion}
\subsection{Summary}
Video streaming is one of the most prominent Internet-based services and is increasingly dominating  the Internet traffic. 
Offering high quality and uninterrupted video streaming service is a complicated process and involves divergent technologies---from video streaming production technologies to techniques for playing them on a widely variety of display devices. With the emergence of cloud services over the past decade, video stream providers predominantly have become reliant on various services offered by the clouds.

In this study, \emph{first}, we explained the workflow and all the technologies involved in streaming a video over the Internet. We provided a bird's-eye-view of how these technologies interact with each other to achieve a high quality of experience for viewers. \emph{Second}, we provided details on each of those technologies and challenges the video streaming researchers and practitioners are encountering in those areas. \emph{Third}, we reviewed the ways various cloud services can be leveraged to cope with the demands of video streaming services. We surveyed and categorized different research works undertaken in this area. 

The main application of cloud services for video streaming can be summarized as: \textbf{(A)} Cloud computational services via VM, containers, or function (\ie serverless computing) paradigms. Computational services can be used for processing video transcoding, video packaging, video encryption, and stream analytics; \textbf{(B)} Cloud network services via Cloud-based Content Delivery Networks (CDN) to reduce video streaming latency, regardless of viewers' geographical location; and \textbf{(C)} Cloud storage services to store video streaming repositories and enable persisting multiple versions of each video to support a wide range of display devices. 

\subsection{Future of Cloud-based Video Streaming Research}
\begin{figure}[htb] 
    \centering
    \includegraphics[width=4in]{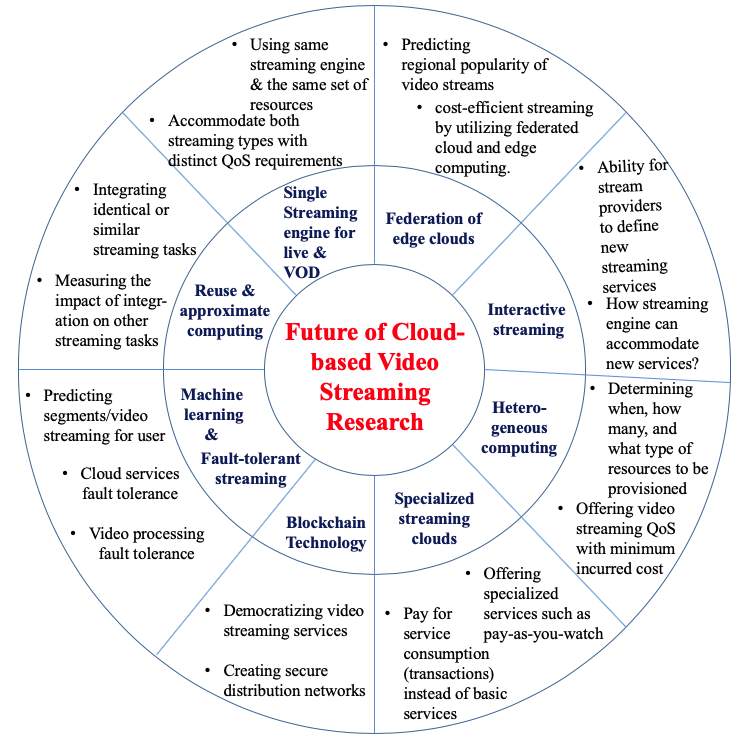}
    \caption{Summary of the future research directions for cloud-based video streaming.}
    \label{fig:future_work}
\end{figure}

Although cloud services have been useful in resolving many technical challenges of video streaming, there are still areas that either remain intact or require further exploration by researchers and practitioners. In the rest of this section, we discuss several of these areas that we believe addressing them will be impactful in the future of video streaming industry. A summary of the future directions are provided in Figure~\ref{fig:future_work}.

\begin{enumerate}

\item \textbf{Interactive video streaming using clouds.}

Interactive streaming is defined as to provide the ability for video stream providers to offer any form of processing, enabled by video stream providers, on the videos being streamed~\cite{hpcc17,hlsaas}. For instance, a video stream provider in e-learning domain may need to offer video stream summarization service to its viewers. Another example can be providing a service that enables viewers to choose sub-titles of their own languages. Although some of these interactions are currently provided by video stream providers\footnote{For instance, YouTube allows viewers to choose their preferred spatial resolution}, there is still not a generic video streaming engine that can be extended dynamically by the video provider to offer new streaming services.

Since there is a wide variety of possible interactions that can be defined on a video stream, it is not feasible to process and store video streams in advance. Instead, they have to be processed upon the viewer's request and in a real-time manner. It is also not possible to process the video streams on viewers' thin-clients (\eg smart-phones), due to energy and compute limitations~\cite{lin17}. Providing such a streaming engine entails answering several researches and implementation problems including: 
\begin{itemize}
 \item How to come up with a video streaming engine that is extensible and a high-level language for users to extend the engine with their desired services and without any major programming effort?
 
 \item How does the streaming engine can accommodate a new service? That implies answering how the streaming engine can learn the execution time distribution of tasks for a user-defined interaction? How can the streaming engine provision cloud resources (\eg VMs or containers) and schedule video streaming tasks on the allocated resources so that the QoE for viewers is guaranteed and the minimum cost is imposed to the stream provider?

\end{itemize}

\item \textbf{Harnessing heterogeneity in cloud services to process video streams.}

Current cloud-based video stream providers predominantly utilize homogeneous cloud services (\eg homogeneous VMs) for video stream processing~\cite{pre_3,pre_5}. However, cloud providers own heterogeneous machines and offer heterogeneous cloud service~\cite{li2018cost}. For instance, they provide heterogeneous VM types, such as GPU, CPU-Optimized, Memory-Optimized, IO-Optimized in Amazon cloud~\cite{li2018cost}. The same is with heterogeneous storage services (\eg SSD and HDD storage services). Heterogeneity may also refer to the reliability of the provisioned services. For instance, cloud providers offer \emph{spot} and \emph{reserved} compute services that cost remarkably lower than the normal compute services~\cite{Kumar2018}.

Traditionally, elastic resource provisioning methods for cloud determine \emph{when} and \emph{how many} resources to be provisioned. By considering heterogeneity, a third dimension emerges: \emph{what type} of resources should be provisioned?
The more specific questions that must be addressed are: how can we harness cloud services to offer video streaming with QoE considerations and with the minimum incurred the cost for the video stream provider? How can we learn the affinity of a user-defined interaction with the heterogeneous services? How can we strike a balance between the incurred cost of heterogeneous services and the performance offered? and finally, how should the heterogeneity of the allocated resources be modified (\ie reconfigured) with respect to the type and rate arriving video streaming requests?

\item \textbf{Specialized Clouds for Video Streaming.}

So far, the idea of cloud has been in the form a centralized facility that can be used for diverse purposes. However, as clouds are evolving to be more service-oriented, their business model is shifting to specific purpose service providers. The reason being service consumers tend to purchase high-level services with certain QoE characteristics (\eg watching video and paying in a pay-as-you-watch manner) as opposed to basic services such as dealing with VMs~\cite{buyya2017manifesto} and pay on an hourly basis. Another example can be purchasing transcoding or transmuxing service with certain startup delay guaranteed. 

In addition, viewers need flexible and diverse subscription services for video streaming services they receive that is not offered under current general purpose cloud services. For instance, some viewers may prefer to pay for video content in a pay-as-you-watch manner and some others may prefer monthly flat rate subscriptions. Offering such detailed facilities entails creating cloud providers dedicated to video streaming services. Creation of such clouds introduces multiple challenges, such as possible connections with basic clouds, specific purpose VM and containers, heterogeneity of underlying hardware among many other challenges. 

\item \textbf{Single streaming engine for both live- and VOD-streaming.}

Live-streams and VODs are structurally similar and also have similar computational demands. However, processing them is not entirely the same. In particular, live-streaming tasks have a hard deadline and have to be dropped if they miss their deadline~\cite{pre_5} whereas VOD tasks have a soft deadline and can tolerate deadline violations. Accordingly, in VOD, upon violating deadlines for a video segment, the deadlines of all segments behind it in the same stream should be updated. That is, VOD tasks have dynamic deadlines which is not the case for live streaming. In addition, there is no historic computational information for live-streaming tasks to predict the execution time of new arriving tasks. This is not the case for VOD tasks.

Based on these differences, video stream providers utilize different video streaming engines and even different resources for processing and providing the services~\cite{pre_4,pre_5}. The question is that how can we have an integrated streaming engine that can accommodate both of the streaming types simultaneously on a single set of allocated resources? How will this affect the scheduling and processing time estimation of the streaming engine?

\item \textbf{Blockchain technology for video streaming.}

The idea of blockchain is to create a secure peer-to-peer communication through a distributed ledger (database). In this system, every network node owns a copy of the blockchain, and every copy includes the full history of all recorded transactions. All nodes must agree on a transaction before it is placed in the blockchain. 

The idea has rapidly got adopted and is being extensively developed in various domains to improve traceability and auditability~\cite{buyya2017manifesto}. We envisage that this technology will have a great impact in the video streaming industry. Some of the applications can be as follows:  
White-listing, which means keeping a list of legitimate publishers or distributors; Identity Management: the ability to perform identity management in a decentralized manner. 

In addition, blockchain technology will grant more control and opportunity to video producers/publishers. In fact, the current video streaming industry is driven by quality expected and algorithms embedded in stream service providers. For instance, ordinary people cannot be publishers on Netflix. Even in the case of YouTube that enables ordinary users to publish content, the search and prioritization algorithms are driven by YouTube and not by the publisher. These limitations are removed in blockchain and publishers have more freedom in exposing their generated content.
The secure distribution network of blockchain can be also useful for current streaming service providers (\eg Netflix). They can use the network to securely maintain multiple versions of videos near viewers and distribute them with low latency and at the meantime reduce their cloud storage costs. 

\item \textbf{Reuse and approximate computing for video streaming.}

Several video streaming viewers may request the same video under the same or dissimilar configurations (\ie processing requirements). Current scheduling methods treat each video streaming request separately. That is, for each video streaming request, they have to be processed separately, even though the same video is being streamed simultaneously for two different viewers. 

To make video streaming cost-efficient, for similar streaming requests (\eg streaming the same video for two users with different resolutions) instead of repeatedly decoding the original video and then re-encoding it, we can reuse processing and decreases the total processing time. For instance, the decoding of a video segment can be done once across all similar requests and then re-encoding is performed separately. In addition to saving cost, this can potentially reduce congestion in scheduling queues by reducing the number of tasks waiting for processing and shortening the overall execution time. While this approach is interesting and there are preliminary efforts to address that (\eg~\cite{chavit18}), it can be challenging as integrating processing of the same video can jeopardize other video streaming tasks to be starved and viewers' QoE is impacted. 

Research studies are needed to be undertaken and address this challenge. One question can be on how to integrate video streaming tasks from different viewers to make a more efficient use of cloud services? How does this approach impact other video streaming tasks? Also, some video streaming tasks might be semantically similar and with some approximation, the result of processing for one request can be used for another request. For instance, two resolutions can be close and compatible. Then, the question is how to identify semantically similar streaming tasks in the system?

\item  \textbf{Machine learning for video processing.}

Video encoding/decoding requires multiple predictions, \eg intra prediction and inter prediction~\cite{zhang2000video}. Machine learning can play an important role to keep improving these predictions to produce smaller size video with the identical quality.

An accurate task execution time estimation can significantly benefit task scheduling and resource provisioning in cloud~\cite{tpds17,ica3pp10}. However, predicting the time is not effortless. 
It is proven that there is an affinity between GOP size and certain video stream processing tasks (\eg transcoding time). However, better estimation can be achieved by using machine learning approaches. 

Deneke \etal~\cite{deneke2014video} utilize a machine learning technique to predict each video segment's transcoding time before scheduling. It shows significantly better load balancing than classical methods. Accordingly, one future research direction in cloud-based video streaming is to use the machine learning techniques to enable unsupervised learning of video streaming tasks for user-defined interactions. The estimation can be different for heterogeneous cloud VMs because various video processing tasks have a different affinity with the heterogeneous VMs.

\item  \textbf{Reliable cloud-based video stream processing}

Reliability of a video streaming service is based on its tolerance against possible failures. Streaming service providers receive services under a certain Service Level Agreement (SLA) with the cloud service provider. SLA explains the availability of services and the latency of accessing them. A video streaming engine translates the SLA terms to its Service Level Objectives (SLO)~\cite{liu2017} and attempts to respect them even in the presence of failures. Failures can be of two types: cloud service (\eg VM or container) failures, and video streaming tasks failures. 
\begin{itemize}
\item \textbf{Cloud service fault tolerance.}
Cloud service (\eg VM) availability is vital for streaming service providers. To maintain good availability, when one server fails, its workloads need to be migrated to another server to keep the streaming service uninterrupted. Service fault tolerance has been widely studied in cloud computing and solutions for that mainly include redundancy of cloud services and data checkpointing~\cite{ Malik_2011, Bala_2012}.  

\item  \textbf{Video processing fault tolerance.}
Some video streaming tasks can fail during processing. Video streaming engines should include policies to cope with the failure of video streaming tasks dispatching to the scheduling queue. The policies can re-dispatch the failed task for VOD streams or ignore it for live-streaming.

\end {itemize}
Currently, there is no failure-aware solution tailored for video streaming processing. Given the specific characteristics of video streaming services, in terms of large data-size, expensive computation, and unique QoE expectations, it will be appropriate to investigate failure-aware solutions for reliable video streaming service.

\item  \textbf{Federation of edge clouds for low-latency video streaming.}

To efficiently serve customer demands around the world, cloud service providers setup datacenters in various geographical locations~\cite{AminiJPDC,aminiAina12}. For example, Netflix utilizes Open Connect~\cite{openconnect18} in numerous geographical locations to minimize latency of video streaming. 

However, existing cloud-based video streaming systems do not fully take advantage of this large distributed system to improve quality and cost of streaming. Mechanisms and policies are required to dynamically coordinate load distribution between the geographically distributed data centers and determine the optimal datacenter to provide streaming service for each video (\eg for storage, processing, or delivery). 

To address this problem, Buyya \etal~\cite{AminiJPDC} advocate the idea of creating the federation of cloud environments. In the context of video streaming, a cost-efficient and low latency streaming can be achieved by federating edge datacenters and take advantage of cached contents or processing power of neighboring edge datacenters. Specifically, solutions are required to stream a video not only from the nearest datacenter (the way CDNs conventionally operate), but also from neighboring edge datacenters. Such solutions should consider the trade-off between the cost of processing a requested video on a local edge datacenter, and getting that from a nearby edge. 

\end{enumerate}

\bibliographystyle{plain}
\bibliography{reference}

\end{document}